\documentclass[12pt]{JHEP3}
\title{Cosmic F- and D-strings}
\author{ Edmund J. Copeland\\ Department of Physics and Astronomy\\
University of Sussex\\
Brighton, BN1 9QJ, UK\\
E-mail: \email{e.j.copeland@sussex.ac.uk}}
\author{ Robert C. Myers\\ Perimeter Institute for Theoretical Physics\\
Waterloo, Ontario N2J 2W9 Canada\\
and\\
Department of Physics\\
University of Waterloo\\
Waterloo, Ontario N2L 3G1 Canada\\
E-mail: \email{rmyers@perimeterinstitute.ca}}
\author{Joseph Polchinski\\ Kavli Institute for Theoretical Physics\\
Santa Barbara, CA\\ 93106-4030, USA\\
E-mail: \email{joep@kitp.ucsb.edu}}
\abstract{
Macroscopic fundamental and Dirichlet strings have several potential
instabilities: breakage, tachyon decays, and confinement by axion domain
walls.  We investigate the conditions under which metastable strings can
exist, and we find that such strings are present in many models.  There are
various possibilities, the most notable being a network of $(p,q)$ strings. 
Cosmic strings give a potentially large window into string physics.}
\keywords{}
\preprint{}

\usepackage{epsfig, cite}
\newcommand{\OL}[1]{ \hspace{1pt}\overline{\hspace{-.5pt}#1
    \hspace{-1.5pt}}\hspace{1.5pt} }
\newcommand{\skyp}[1]{}

\newcommand{\klmt}{\mbox{K\hspace{-7.6pt}KLM\hspace{-9.35pt}MT}\ }
\newcommand{\KLMT}{\mbox{K\hspace{-12pt}KLM\hspace{-15.5pt}MT}\ }
\def\gsim{\stackrel{>}{{_\sim}}}
\def\lsim{\stackrel{<}{{_\sim}}}

\begin{document}
\section{Introduction}

Before the `second superstring revolution' there appeared to be a clear
distinction between fundamental strings and cosmic strings.   Fundamental
strings were believed to have tensions $\mu$ close to the Planck scale. In
perturbative heterotic string theory, for example, $G\mu = 
\alpha_{\rm GUT} / 16\pi \gsim 10^{-3}$, whereas the isotropy of the
cosmic microwave background implied (even before COBE) that any string of
cosmic size must have $G\mu \lsim 10^{-5}$~\cite{cosmicstrings}.  Thus any
cosmic strings would have had to arise in the low energy effective  field
theory, as magnetic or electric flux tubes.

In principle, fundamental strings could have been produced in the early
universe and then grown to macroscopic size with the expansion of the
universe.  Inflation provides a simple explanation for the absence of
cosmic fundamental strings of such high tension.  However, even aside  from
inflation it was noted in ref.~\cite{witten} that there are effects that
would prevent the appearance of cosmic fundamental strings. Macroscopic
type I strings break up on a stringy time scale into short open strings,
and so would never form.  Macroscopic heterotic strings always appear as
boundaries of axion domain walls, whose tension would force the strings to
collapse rather than grow to cosmic  scales~\cite{VE}. At the time of
ref.~\cite{witten} no instability of long type II  strings was known, but
it is now clear that NS5-brane instantons~\cite{ns5} (in combination with
supersymmetry breaking to lift the zero modes) will produce an axion
potential and so lead to domain walls.

Today the situation in string theory is much richer.  First, many new
one-dimensional objects are known: in addition to the fundamental
F-strings, there are D-strings, as well as  higher dimensional D-, NS-,
M-branes that are partially wrapped on compact cycles so that only one
noncompact dimension remains.  Second, the possibility of large compact
dimensions~\cite{lcd} and large warp factors~\cite{RS} allows much lower
tensions for these strings.  Third, the various string-string and
string-field dualities relate these  objects to each other, and to the
field-theoretic flux tubes, so that they are actually the same object as
it appears in different parts of parameter space.  Thus it is important to
revisit this subject, and ask whether some of these strings may be
cosmologically interesting.

Indeed, it has been argued by Jones, Sarangi, and Tye and by Stoica
and Tye~\cite{tye} that D-brane-antibrane inflation~\cite{bab} leads
to the copious production of lower-dimensional D-branes that are
one-dimensional in the noncompact directions.  This is a special case
of the production of strings in hybrid inflation~\cite{hyb}. 
Refs.~\cite{tye} also make the important observation that
zero-dimensional defects (monopoles) and two-dimensional defects
(domain walls) are not produced; either of these would have led to
severe difficulties. 

It is necessary then to investigate the stability of possible cosmic strings
against the processes noted above.  A naive extrapolation of the results of
ref.~\cite{witten}  would suggest that all BPS strings are confined by
domain walls, and that all  non-BPS strings are unstable to breakage or
tachyon decay.  We will find  that the situation is more interesting, and
that stable strings exist in certain classes of models but not in others.

In \S2 we investigate this subject, and identify conditions under which long
strings can be at least metastable.  Our focus is on type I/II theories,
though the same principles will hold in heterotic and M theory
compactifications.  In \S3 we apply our conditions in the string theory
inflation model of Kachru, Kallosh, Linde, Maldacena, McAllister, and Trivedi
(\mbox{K\hspace{-7.6pt}KLM\hspace{-9.35pt}MT})~\cite{KKLMMT}, which is based
on the stabilization of all moduli in the warped IIB framework of
ref.~\cite{KKLT}.  We find that the nature of the  cosmic strings in this
model depends on precisely how the Standard Model fields and the moduli
stabilization are introduced. We identify three possibilities: (a) no
strings; (b) D1-branes only (or fundamental strings only); (c) $(p,q)$
strings --- bound states of
$p$ fundamental strings and $q$ D-strings for relatively prime 
$(p,q)$ --- with an upper bound on $p$.  In \S4 we briefly discuss large
compact dimension models without large warp factors, and find a similar
range of possibilities. In \S5 we discuss the observational signatures of
these cosmic strings.  Although the various bounds currently are all in the
area of $G\mu
\lsim 10^{-6}$, there are future observations that will reach many orders of
magnitude further in $G\mu$.  With $(p,q)$ strings there are more
complicated string networks than when only one type of string is present. 
This may enhance the signatures for these strings, and possibly place strong
constraints on models of these types.

While we were completing this work we learned of papers on related
subjects by Dvali, Kallosh and Van Proeyen~\cite{DKVP} and by Dvali and
Vilenkin~\cite{DVnew}.

\section{Stability Conditions}

\subsection{Breakage}

\subsubsection{Breakage on an ${M}^4$-filling brane}

The prototype example of string breakage is the type I string.  On a  long
type I string there is a constant rate per unit length for the string  to
break by formation of a pair of endpoints~\cite{break}.
\EPSFIGURE[h]
{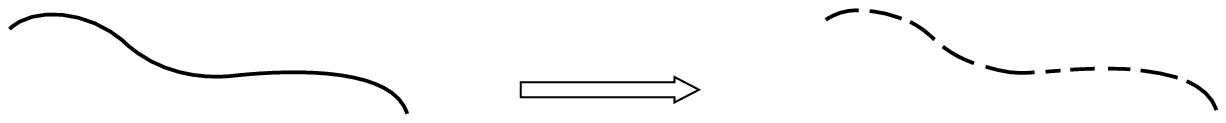}
{A long open string converts to short open strings on a stringy 
time scale.}
By this process the string will rapidly convert to ordinary quanta as in figure~1.  The
modern interpretation is that these endpoints are attached to a
spacetime-filling D9-brane.  Similarly, the fundamental type II string
(F-string) can end on any D-brane, and so in any type II model with
D$p$-branes that fill the noncompact $M^4$ dimensions there will be an
amplitude for the string to break.  Of course, for $p<9$ the D-brane does
not fill the {\it compact} dimensions, and there is the possibility that
the breaking is suppressed due to transverse separation between the string
and the D-brane.  This will allow metastable strings to exist in some
models.

\EPSFIGURE[h]
{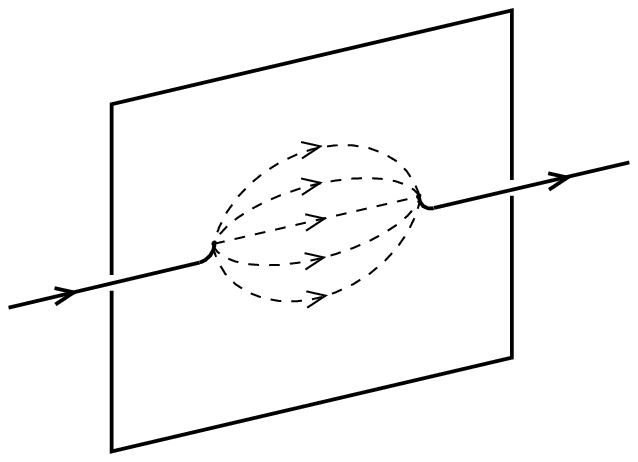}
{When a string ends on a brane, flux lines in the brane still 
connect
the endpoints.}

By $S$-duality from F1/D3, a D-string can end on a
D3-brane~\cite{andy,schwarz} and so will be unstable in compactifications
with D3-branes.  A  D-string
can also end on a D5-brane, but actually this does not represent an
instability.  When an F- or D-string ends on a D3-brane, the endpoints 
are electric or magnetic sources on the brane.
The endpoint pair that is
produced when a string breaks are still connected by flux lines in the
brane --- one can think of the string as dissolving into the brane and
becoming flux. For a D-string oriented in the 1-direction, the flux is
one unit of integrated $F_{23}$, which can lower its energy toward zero
by spreading in the 23-plane as in figure~2.  Similarly an F-string dissolving in any
D-brane becomes one unit of $F_{01}$ which can again spread.   For a
D-string dissolving in a D5-brane however, there is one unit of $F 
\wedge F$ in the four transverse directions, and the energy of this flux
is not lowered by spreading out --- this is the scale invariance of the 
instanton action.  In other words, there is still a nonzero energy per
unit  length, which is not surprising given that the parallel D1/D5 system
already saturates the BPS bound.\footnote{The exact scale invariance will 
be broken as a consequence of supersymmetry breaking, but there will still
be a positive minimum to the string tension.}  So from the four-dimensional
point of view  the string does not break, and we might say that while a
D-string can {\it attach} to a D5-brane it cannot {\it break} on one. 
Similarly it  actually costs energy for a D-string to attach to a D7-brane,
and so again the  string cannot break.

As another example, a string which arises from a D$3$-brane wrapped on a
$2$-cycle can end on a D$5$-brane which wraps the same cycle and fills the
noncompact dimensions.  Other cases can be obtained from these by $S$- and
$T$-duality.  

In summary, stability requires that there 
be no $M^4$-filling branes on which the string can end, or that the decay
be suppressed by transverse separation.
We will also encounter a related instability, the tachyonic decay of an
unstable D-brane.  This is similar to breakage in that the D-brane turns
into ordinary quanta on a stringy time scale by decays everywhere along its
length.  An unstable D-brane in superstring theory is constructed from a
D-$\OL{\rm D}$ pair~(for a review see~ref.~\cite{nonbps}), and the tachyon
is an open string  stretching between them. As with breakage, this decay
can also be suppressed by transverse separation, in this case of the D and
the $\OL{\rm D}$.

\subsubsection{Breakage on a `baryon'}

The $U(1)$ gauge field on the $M^4$-filling brane plays an essential role in
allowing the strings to break.  Strings couple to forms, for example
\begin{equation}
\int_{\rm F_1} B_{(2)}\ ,\qquad \int_{\rm D_1} C_{(2)} \label{stringact}
\end{equation} 
for an F-string or D-string respectively,  where the
integrals run over the string world-volumes.\footnote{Our notation is that
$C_{(q)}$, $B_{(2)}$, and $A_{(1)}$ are the R-R, NS-NS, and brane gauge
potentials, and $F_{(q+1)}$, $H_{(3)}$, and $G_{(2)}$ the corresponding
field strengths.  A tilde on a field strength denotes the inclusion of
Chern-Simons terms.  A four-dimensional form obtained by integrating a
higher-rank form over a compact submanifold is denoted by square brackets,
e.g. $C_{[q]}$.} For a world-volume without boundary, these actions are
invariant under the form gauge transformations 
\begin{equation}
\delta B_{(2)} =
d\lambda_{(1)}\ ,\quad\delta C_{(2)} = d\chi_{(1)}\ , 
\end{equation}
but when the string
breaks there is a surface term in the transformation.   The point is that
the form field also couples to the $U(1)$ flux, and one sees from figure~2
that this provides the necessary continuity.  The relevant terms in the
brane action are 
\begin{equation}
S' =\int_{\rm D3}  \frac{1}{2} |G_{(2)} + B_{(2)}|^2 + G_{(2)} \wedge C_{(2)}\ .
 \label{d3act}
\end{equation}
Here $G_{(2)} = d A_{(1)}$ is the gauge field strength on the brane, and we have included its Chern-Simons couplings to the forms.
We take the example of the D3-brane, and normalizations are simplified for
clarity.  The variation of $S'$ under the form transformations is
\begin{equation}
\delta S' = -\int_{\rm D3}  \lambda_{(1)} d {*} ( G_{(2)} + B_{(2)} ) + \chi_{(1)} dG_{(2)}\ . \label{spvar}
\end{equation}
 The F1-brane endpoint is an electric source for $G_{(2)} + B_{(2)}$, and the D-brane endpoint is a magnetic source for $G_{(2)}$, each giving a delta function in one of the terms in the variation~(\ref{spvar}).  Thus the variation of the actions~(\ref{stringact})
and~(\ref{d3act}) cancel when a string breaks on a brane.
  The apparent
asymmetry between the forms $B_{(2)}$ and $C_{(2)}$ in the
action~(\ref{d3act}) can be reversed by an electric-magnetic duality
transformation on the brane.

Type II theories also have $U(1)$ gauge fields arising from the bulk RR
fields, and these can play the same role.  Consider the IIB theory, for
example.  The four-form potential with one index along $M^4$ and three in
the compact directions gives rise to a four-dimensional gauge field for each
three-cycle ${\cal K}_3$
\begin{equation}
C_{[1]} = \int_{{\cal K}_3} C_{(4)}.
\end{equation}
The five-form field strength includes a Chern-Simons term $\tilde F_{(5)} =
d C_{(4)} + B_{(2)} \wedge F_{(3)}$.  One is generally interested in
compactifications in which some of the background three-form fluxes are
nonvanishing.  If, for example, 
\begin{equation}
\int_{{\cal K}_3} F_{(3)} = M\ , \label{3flux}
\end{equation}
then the effective two-dimensional field strength is 
\begin{equation}
\tilde F_{[2]} = \int_{{\cal K}_3} \tilde F_{(5)} = dC_{[1]} + M B_{(2)} \ .
\end{equation}
Comparing with the first term in the action~(\ref{d3act}), we see that we have the necessary
coupling to allow the F1-brane to break; the role of the integer $M$ will be
seen below.  

To complete the picture we need the mechanism by which the string breaks. 
Consider a D3-brane wrapped on the same cycle ${\cal K}_3$.  This is a
localized particle in four dimensions, which we refer to loosely as a
baryon, because this is the role it plays in gauge/string duality~\cite{baryo}. The gauge
field equation on this brane's world-volume is $d * G_{(2)} = F_{(3)}$, again
following from the action~(\ref{d3act}), and this is inconsistent with the
flux~(\ref{3flux}) unless $M$ F1-branes end on the baryon.  We conclude that
precisely if $M = 1$ the F1-brane can break by pair production of baryons. 
If $M \geq 2$ then instead the baryon is a vertex at which $M$ F-strings
meet.

More generally, if  
\begin{equation}
\int_{{\cal K}_3}F_{(3)} = M\ , \quad \int_{{\cal K}_3} H_{(3)} = M'
\end{equation}
then the baryon is a vertex at which $M$ F1-branes and $M'$ D1-branes end.

\subsection{Axion domain walls}

Any BPS $p$-brane (that is, the infinite flat brane is a BPS state)
must couple to a $(p+1)$-form field $C_{(p+1)}$ (we are using the notation
for an R-R brane, but the same applies to NS-NS and M theory branes).  This 
gives the repulsive force that offsets the gravitational and
dilatonic attraction between a  pair of parallel branes.  Let the brane be
wrapped on a
$(p-1)$-cycle
${{\cal K}_{p-1}}$, thus leaving a string in the noncompact directions.  In
four dimensions there will be a two-form potential
\begin{equation}
C_{[2]} = \int_{{\cal K}_{p-1}} C_{(p+1)}\ ,
\end{equation}
which couples to the world-volume of the string.  In the  four-dimensional
theory this is dual to an axion field $\phi$,
\begin{equation}
dC_{[2]} = *_4 d\phi + \mbox{source terms}\ .
\end{equation}
The string is an electric source for $C_{[2]}$ and so a magnetic,
topological, source for $\phi$.  That is, the axion field (appropriately
normalized) changes by $2\pi$ in going around the string:
\begin{equation}
\oint_C dx \cdot \partial \phi = 2\pi  \label{circ}
\end{equation}
on any contour $C$ encircling the string.  
  
  Now consider a Euclidean
$(6-p)$-brane instanton, which couples  magnetically to $C_{(p+1)}$,
wrapping a $(7-p)$-cycle ${\cal K}_{7-p}$ that intersects ${\cal K}_{p-1}$ once.  A magnetic source
for $C_{(p+1)}$ is an electric source for $\phi$, and so the instanton
amplitude is proportional to $e^{i\phi}$.  Since all supersymmetries are
ultimately broken, the fermion zero modes in the instanton amplitude are
lifted, and this produces a periodic potential for $\phi$.  From
eq.~(\ref{circ}) it follows that $\phi$ cannot sit in the minimum of this
potential everywhere as we encircle the string ---  there is a kink where
it changes by $2\pi$ and passes over the maximum of the potential.  Since
the kink in $\phi$ intersects any contour $C$ that circles the string, it  defines a domain wall ending on the string.  Unless the domain wall tension is exceedingly small this will  cause the strings to rapidly disappear~\cite{VE}.
  
\subsection{Two puzzles}

\subsubsection{The first puzzle}

A string cannot {both} break and bound a domain wall, because the boundary 
of a boundary is zero.  What happens if both instabilities appear to be
present?  Consider the case of a D1-brane which can end on a D3-brane 
and which also couples to a massless four-dimensional field $C_{(2)}$.  The effective four-dimensional action is
 \begin{equation}
\frac{1}{2}\int |dC_{(2)}|^2 + |G_{(2)} |^2 + G_{(2)} \wedge C_{(2)}\ .
\end{equation}
We are writing explicitly only the terms that govern the local dynamics, not the source
terms.
For a higher dimensional wrapped D-brane one just replaces $C_{(2)}$ by $C_{[2]}$, and for an F-string by $B_{(2)}$ (after a D3 world-volume electric/magnetic duality).
The effective four-dimensional field equation for the massless field $C_{(2)}$ is
\begin{equation}
d {*_4} d C_{(2)} = G_{(2)}\ , \label{afel}
\end{equation}
which implies that the axion field must be defined
\begin{equation}
d C_{(2)} = {*_4} (d\phi + A_{(1)}) \ .
\end{equation}
It follows that $\phi$ transforms nonlinearly under gauge 
transformations of the brane gauge field $A_{(1)}$, 
 \begin{equation}
 \delta A_{(1)} = d\lambda\ ,\quad \delta\phi = -\lambda\ , \label{nlhiggs}
 \end{equation}
 and so the $U(1)$ on
the brane is Higgsed.  There is no domain wall because the
axion is removed by the Higgs mechanism.\footnote{The instanton amplitude
$e^{i\phi}$ would seem to violate gauge invariance,  but this is balanced
by the production of charged strings.  Consider for  example a D1 string
which can end on a D3-brane and whose axion couples to a Euclidean D5
instanton.  The instanton and D3-brane intersect in a  spacetime point,
and so the string creation process is essentially the same as  for a
D0-brane passing through a D8-brane~\cite{create}.}  

Let us look in more detail at the external field of a D-string running in
the 1-direction, again in the effective four-dimensional picture.  The nonzero field components are $C_{01}$ and
$G_{23}$.  The field equation for $A_{(1)}$,
 \begin{equation}
 d(*G_{(2)} + C_{(2)}) = 0
\end{equation}
implies that $C_{01} - G_{23}$ is constant in the transverse directions, and so vanishes.  The field equation~(\ref{afel}) with source term from the D-string
then implies that
\begin{equation}
\partial_\perp^2 C_{01} = C_{01} + \delta^2(x_\perp)\ .
\end{equation}
It follows that the R-R field of the string falls exponentially at infinity
(the mass is 1 because the constants have been dropped).  The R--R field of the D-string is screened, so there is no obstruction to the D-string breaking.  The D-string dissolves in the D3-brane to become a tube of $G_{23}$ flux; the total flux in this tube is equal and opposite to the screening flux, and so the $G_{23}$ field dissolves into ordinary quanta.

\subsubsection{The second puzzle}

We have concluded that there is no stable D-string.  However, there is a spontaneously broken $U(1)$ gauge symmetry~(\ref{nlhiggs}), and so one might have expected a topologically stable string associated with this broken symmetry~\cite{stabl}.  Such a string would involve degrees of freedom beyond those that we are considering, namely the `radial' Higgs field associated with the angular part $\phi$, and so a higher energy scale; whether such strings exist depends on the topology of this higher-energy field space.  

Unlike D-strings, such `$\phi$-strings' would not generically be produced in brane-antibrane inflation.  Ref.~\cite{tye} shows that the production of strings can be understood by applying the Kibble argument~\cite{TWB} to the vacuum manifold for the brane-antibrane tachyon, as in K-theory~\cite{nonbps,ktheory}.  Add in a D3-$\OL{\rm D3}$ pair, such as would be present during inflation.
If we first ignore the Higgsing by $\phi$, the vacuum manifold is
\begin{equation}
\frac{U(2)\times U(1)}{U(1)\times U(1)} = S^3\ . \label{2to1}
\end{equation}
The breaking is the same as the Weinberg-Salam model, $SU(2) \times U(1) \to U(1)$.  The vacuum manifold is simply connected, so there is no stable D-string, reflecting the fact that it can break on the surviving D3-brane.

Now include the the nonlinear Higgsing~(\ref{nlhiggs}).  This breaks one of the denominator $U(1)$'s.  However, the important point is that this $U(1)$ is broken even before the brane annihilation, by the same nonlinear mechanism: one of the numerator $U(1)$'s is broken as well.  Thus the
vacuum manifold for D3-$\OL{\rm D3}$ annihilation is still $U(2)/U(1) = S^3$ and so the Kibble mechanism does not lead to strings.  Any strings associated with this already-broken symmetry will have been diluted away during inflation, and the nonlinear Higgs field will be ordered over long distances.  Of course, if the number of $e$-foldings is sufficiently small, or if additional phase transitions take place after inflation, then there may be interesting strings beyond those that we consider.

\section{The \KLMT model}

\subsection{Generalities}

We now apply the above analysis to the \klmt model~\cite{KKLMMT}.  This is
presently the most well-developed model of inflation in string
theory, being based on a framework in which all moduli are
stabilized~\cite{KKLT}.  Although many open issues remain, some of which
will be encountered below, this provides an excellent test case for cosmic
strings.

The \klmt model is based on IIB string theory on a Calabi-Yau manifold.
\EPSFIGURE[h]{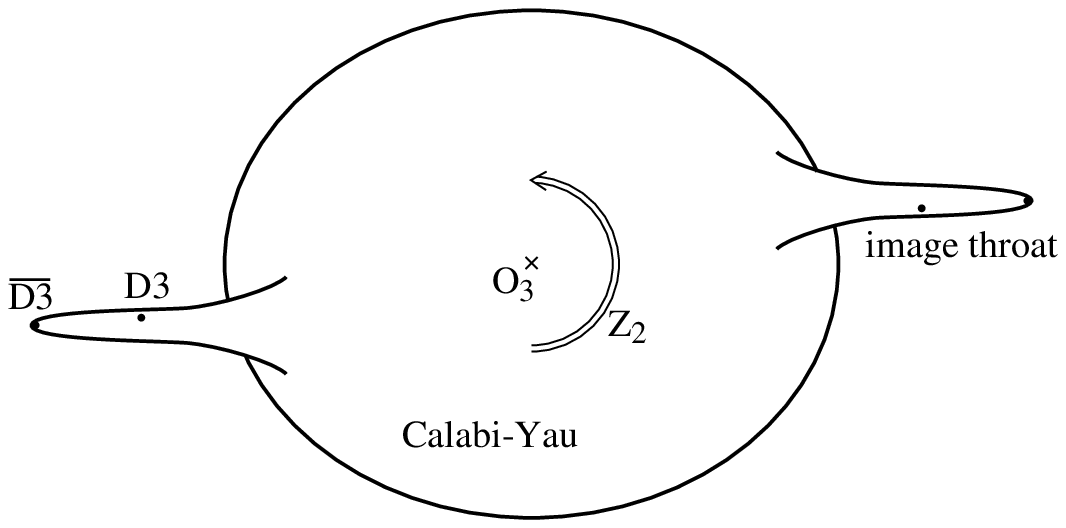}
{Schematic picture of the \klmt geometry: a warped Calabi-Yau
manifold with throats, identified under a ${\bf Z}_2$ orientifold.
}
This is orientifolded by a ${\bf Z}_2$ symmetry that has
isolated fixed points, which become O3-planes.\footnote{We focus on this
special case of the more general F theory construction, but as we note
below the results in the latter case are much the same.}  The spacetime
metric is warped,
\begin{equation}
ds^2 = e^{2A(x_\perp) }\eta_{\mu\nu} dx^\mu dx^\nu + ds_\perp^2\ .
\end{equation}
The inflaton is the separation between a D3-brane and an anti-D3-brane,
whose annihilation leads to reheating.  The annihilation occurs in a region
(throat) of large gravitational redshift, ${\rm min}\{ e^{A(x_\perp)} \} = e^{A_0}
\ll 1$, where
$e^{A(x_\perp)}$ is normalized to be $O(1)$ in the bulk of the Calabi-Yau.
Note that in figure~3 the covering manifold has two
throats which are identified under the ${\bf Z}_2$ rather than a single
throat identified with itself; this is equivalent to the statement that
there is no O3-plane in the throat.

The redshift in the throat plays
a key role: both the inflationary scale and the scale of string tension, as
measured by a ten-dimensional inertial observer, are set by string
physics and are close to the four-dimensional Planck scale.
The corresponding energy scales as measured by a four-dimensional physicist
are then suppressed by a factor of $e^{A_0}$.

As discussed in ref.~\cite{tye}, one expects copious
production only for objects that are one-dimensional in the noncompact
dimensions and lie entirely within the region of reheating.  The
obvious candidates are then the F1-brane (fundamental IIB string) and
D1-brane, localized in the throat.  There is also another possibility,
which however can be quickly disposed of.  In the \klmt model, the throat
is a Klebanov-Strassler geometry~\cite{KS}, whose cross section  is
topologically $S^2 \times S^3$.  A D3-brane wrapped on the $S^2$ also
gives a string in four dimensions.  However, the $S^2$ is topologically
trivial, it collapses to a point at the end of the throat~\cite{KS}, 
and so this string can break rapidly.

The D1-branes can be regarded as topological defects in the tachyon field
that describes D3-$\OL{\mbox{D3}}$ annihilation~\cite{nonbps,ktheory}, and
so these will be produced by the Kibble mechanism~\cite{TWB}.  The F1-branes do not
have a classical description in these same variables, but in an $S$-dual
description they are topological defects and so must be produced in the
same way.  Of course, only one of the $S$-dual descriptions can be
quantitatively valid, and if the string coupling is of order one then
neither is.  However, the Kibble argument depends only on causality and so
it is probably valid for both kinds of string in all regimes (it has also been argued in ref.~\cite{DV1} that fundamental strings will be created).
All strings created at the end of inflation are at the bottom of the
inflationary throat, and they remain there because they are in a deep
potential well: their effective four-dimensional tension $\mu$ depends on the
warp factor at their location and their ten-dimensional tension $\bar\mu$ as
\begin{equation}
\mu = e^{2A(x_\perp) }\bar\mu \ .  \label{redten}
\end{equation}

Given both F1-branes and D1-branes, there will also be $(p,q)$ strings for
relatively prime integers $p$ and $q$.  These were found in
ref.~\cite{pq} using the $SL(2,{\bf Z})$ duality of the IIB
superstring theory, and are now interpreted~\cite{joed,bound} as bound
states of $p$ F1-branes and $q$ D1-branes.  Their tension in the
ten-dimensional theory is
\begin{equation}
\bar \mu_{p,q} = \frac{1}{2\pi\alpha'} \sqrt{ (p - C q)^2 + e^{-2\Phi}
q^2 }\ ,
\end{equation}
where $C$ is the Ramond-Ramond scalar (normalized to periodicity~1) and
$\Phi$ is the dilaton ($e^\Phi = g_{\rm s}$), both evaluated at the location
of the string. In the \klmt model the values of these fields are fixed in
terms of discrete three-form fluxes.

It remains to discuss the stability of these strings.  From the previous
section, it is essential to know what branes remain in the theory
{after} inflation, upon which the
$(p,q)$ strings might break.  In the \klmt model there must be such
branes for two reasons: (a) there must be the branes on
which the Standard Model fields live, and (b)
the moduli stabilization in this model involves one or more  anti-D3-branes
located in a throat (these could possibly be the same as the Standard Model
branes).  

\subsection{Scenarios}

As we will explain, the only branes that are relevant for the stability of
the cosmic strings are those that are located in, or intersect, the
inflationary throat.  We thus consider the various possibilities.
 
\subsubsection{No branes in the throat}

Here we consider the case that the Standard Model and stabilizing branes
are located outside the inflationary throat.  Let us first ask
whether the $(p,q)$ strings are BPS.   In fact they are not.  The easiest
way to see this is to consider the special case of the $T^6/{\bf Z}_2$
orientifold, which is
$T$-dual to the type-I theory.  The D1-branes become type I D7-branes, which
are non-BPS~\cite{ktheory}.  The type I  orientation reversal turns a IIB
D7-brane into a $\OL{\rm D7}$-brane, so the type I D7-brane is actually a
bound state of a IIB D7-$\OL{\rm D7}$ pair.  The F1-branes
become type I fundamental strings, which are again non-BPS.  Another way
to understand this is through the orientifold projection, under which the
forms $B_{\mu\nu}$ and $C_{\mu\nu}$ that couple to the F1- and D1-branes
are odd~\cite{GKP}.  Thus the zero modes of these fields are removed, and
do not appear in the massless spectrum.\footnote{This will also be true in
the more general F theory constructions. The zero modes of
$B_{\mu\nu}$ and $C_{\mu\nu}$ are inconsistent with the nontrivial
$SL(2,{\bf Z})$ holonomy of the 7-branes.}
 
The orientifold projection in the \klmt model turns a D1-brane into an
anti-D1-brane, which is to say it reverses the orientation.  The
four-dimensional D-string is thus in this construction a D1-$\OL{\rm
D1}$ bound state, a potentially unstable
configuration. We have noted above that the D1 is confined to the tip of
the inflationary potential.  The key point is then that the image $\OL{\rm
D1}$ is in the image throat.  The length of each throat is somewhat
greater than unity in string units, so the D1-$\OL{\rm D1}$ strings are
stretched and nontachyonic.  

Even without a tachyon it is possible for the D1-string
to fluctuate into the other well and annihilate with the $\OL{\rm D1}$.  
In order for the D-string to be an interesting cosmic string the decay rate
must be suppressed from its natural string scale, of order $\alpha'^{-1}$
per unit string length and time, to a rate of order the cosmic scale $H^2$
or less, a suppression of roughly $10^{50}$ in time scale and so $10^{100}$
in rate per unit length.  In fact the suppression is much greater than this,
due to the warping of the compact space.  The decay proceeds through the
appearance of a hole in the Euclidean world-sheet, in which the D1 and
$\OL{\rm D1}$ have annihilated.  At the edge of this hole the D1-brane
crosses over the to the image throat and annihilates with the $\OL{\rm D1}$.
The rate is given by the usual Schwinger expression for this instanton,
\begin{equation}
e^{-B}\ , \quad B = \pi \sigma^2 / \rho\ . \label{schwing}
\end{equation}
Here $\rho$ is the action per unit area, {\it i.e.} the tension
$e^{2A_0} /2\pi\alpha' g_{\rm s}$.  The parameter $\sigma$ is the action per
unit length for the boundary of the hole.  Since the D1-brane must pass
through the unwarped bulk of the Calabi-Yau, this is of order $R/2\pi\alpha'
g_{\rm s}$ where $R$ is the distance between the throat and its image.  The
dominant factor in $B$ is from the warping,
\begin{equation}
B \sim e^{-2A_0}  \sim 10^8\ , \label{bigb}
\end{equation}
where the numerical value is taken from ref.~\cite{KKLMMT} and will be
discussed further in \S5.  The warping suppresses the decay by
the impressive factor $e^{-B}$, even though the D1 and its image are only a
few string units apart!\footnote{The radial metric in the throat goes as $dr^2/r^2$, so the length is only logarithmic in $e^{A_0} \sim r_{\rm min}/r_{\rm max}$.}  The decay of the F-string and all the other
$(p,q)$ strings  are similarly stabilized, because they involve world-sheets
that stretch from one throat to its image, through the unwarped region.

We can now see why the branes outside the throat are irrelevant.  In order
for a string to break on one of these, it must fluctuate out of the throat,
which again costs the suppression~(\ref{bigb}).

Finally, let us consider decay via baryon pair production.  The only
relevant baryons are the D3-branes that wrap the $S^3$ in the
Klebanov-Strassler throat.  All other three-cycles pass through the bulk of
the Calabi-Yau, and so the masses of the corresponding baryon are at the
string scale, unsuppressed by the warp factor.  For these the pair
production rate is then suppressed by the same factor~(\ref{bigb}).  

The $S^3$ in the throat carries $M$ units of $F_{(3)}$~\cite{KS,GKP}.  It follows
that if $M=1$ the F-strings are unstable to baryon production.  More
generally, a baryon allows a $(p,q)$ string to decay to a $(p-M,q)$ string,
and so the stable strings have $|p| \leq M/2$.

In the \klmt model there is a lower bound on $M$.  During inflation there
is at least one $\OL{\rm D3}$ in the inflationary throat.  It
is shown in ref.~\cite{KPV} that this is classically unstable unless 
\begin{equation}
M \gsim 12\ .\label{mbound}
\end{equation}
  Thus the baryon-mediated decay
will not destabilize the $(p,q)$ strings with small $p$, and there is still
a rich spectrum.  

The numerical value of the tunnelling rate is again given
by eq.~(\ref{schwing}), where now $\sigma$ is the baryon mass
\begin{equation}
\sigma = \OL T_{\rm D3} V_{S^3}\ , \quad \OL T_{\rm D3} = \frac{1}{(2\pi)^3 \alpha'^2 g_{\rm s}}\ ,
\end{equation}
and
\begin{equation}
\rho = \bar\mu_{p,q} - \bar\mu_{p-M,q}\ .
\end{equation}
Obtaining the volume of the $S^3$ from eq.~(65) of ref.~\cite{HKO}, one
obtains for $(p,q) \to (p-M,q)$ with $p > M/2$ a decay rate  of order
\begin{equation}
e^{-B}  \ ,\quad B \sim 0.2 \times \frac{q M^2}{2p - M} \ ;
\end{equation}
we have taken $g_{\rm s} \ll 1$ to simplify the expression.
There is no suppression from the warp factor because the decay takes place
entirely in the throat, and the decay will be rapid on a
cosmological time scale  $(B < 200)$ for a wide range of parameters.  For example, if $q=1$ and $2p - M = O(M)$, the decay is rapid for $M < 10^3$.
The baryon is a
`bead' at which a $(p,q)$ string and a $(p-M,q)$ string join, and it will
accelerate rapidly in the direction of the higher-tension string.

Incidentally, a process similar to the pair production of baryons is the
decay of the inflationary $\OL{\rm D3}$ by an NS5-brane instanton wrapped on
the $S^3$~\cite{KPV,FLW}.  This is negligible on cosmological time scales unless
$M$ is very close to the bound~(\ref{mbound}).

In summary, in this case we obtain $(p,q)$ strings for relatively prime
$p,q$ with $|p| \leq M/2$.

\subsubsection{Stabilizing $\OL{\rm D3}$-branes in the throat}

Now suppose that the stabilizing $\OL{\rm D3}$-brane(s) are in the same
throat in which inflation occurs.  In other words, inflation is driven by 
$N$ D3-branes and $\Delta+N$ $\OL{\rm D3}$-branes for some $\Delta>0$, so that after
annihilation $\Delta$ $\OL{\rm D3}$-branes remain.  In this case the F-strings
and D-strings can both break, as can all the $(p,q)$ strings, and there
are no cosmic strings.  In terms of K theory, the tachyon
vacuum manifold is
\begin{equation}
\frac{U(N)\times U(\Delta+N)}{U(N)\times U(\Delta)}\ .
\end{equation}
This contains a $U(1)$ and supports D-string vortices only if the
number $\Delta$ of residual $\OL{\rm D3}$-branes vanishes.

The various scenarios that we are exploring must satisfy a number of
nontrivial constraints, including stability of the weak scale,
stability of the moduli during inflation, and sufficient
reheating~\cite{wip}.  We will not attempt to discuss these in detail
here, but it is worth noting that the present scenario has a serious,
probably fatal, problem with reheating.  The point is that there is a
$U(1)$ gauge field on the stabilizing $\OL{\rm D3}$.  This is the only
massless degree of freedom in the inflationary throat, and so the only one
that couples directly to the inflationary fields.  Then almost all of the
energy at reheating goes into these
$U(1)$ gauge bosons rather than into the Standard Model fields.

Of course, many of the specific features of the \klmt model may be
absent in more general constructions.  The role of the $\OL{\rm D3}$-branes
is to break supersymmetry, raising the supersymmetric AdS vacuum to a  state
of approximately zero cosmological constant~\cite{KKLT}.  It seems likely
that there are other dynamical mechanisms that would accomplish the same
thing,\footnote{We thank S. Kachru and E. Silverstein for comments on
this point.} and that there need not be an associated $U(1)$ field in
general.  This eliminates the immediate problem with reheating (though this
is still a nontrivial issue to explore).  However, the alternate stabilizing mechanism no longer mediates breaking of $(p,q)$ strings, as this also requires a
$U(1)$ gauge field as in figure~2.  Thus we would obtain the same strings as
in the no-brane case. 

\subsubsection{Standard model branes in the throat}

Now let us consider the Standard Model branes.  In the \klmt model it is
natural to introduce D3-branes and/or $\OL{\rm D3}$-branes, as well as
D7-branes in the F theory constructions.  Ref.~\cite{AIQU} gives
constructions of the Standard Model in terms of local configurations with
these ingredients.  These local constructions can then be incorporated into
the \klmt model.  In order to reduce the supersymmetry of the low energy
theory from ${\cal N} = 4$ to
${\cal N }= 1$ or ${\cal N} = 0$ it is necessary that the D3-branes (or
$\OL{\rm D3}$-branes) be fixed at an appropriate singularity, for example an
orbifold fixed point.  The D7-branes are needed for anomaly cancellation, and
they must also pass through this singularity.  In some cases (certain
non-Abelian orbifolds) the D7-branes are not needed~\cite{AIQU,BJL}.

Consider first the case that the 3-branes and the singularity are outside the
inflationary throat but at least one of the D7-branes passes through the
bottom of the throat.  By the general discussion of breakage, the F1-branes
will now be able to break but the D1-branes will not, and so the only cosmic
string is the D-string.  However, there is a caveat.  The breaking of the
F-strings requires the $U(1)$ gauge field on the D7-brane.  Since the
D7-branes are extended, their dynamics depends on the physics not only in the
throat but in the full Calabi-Yau.  If the $U(1)$ is confined in some way,
then again a larger set of $(p,q)$ strings will be stable.  This may be
the case, because in F theory compactifications there is not in general a
$U(1)$ gauge field locally associated to each 7-brane~\cite{ftheory}; some of
the $U(1)$'s are frozen out by the global dynamics.

Now consider the case that the Standard Model 3-branes and the associated
singularity are located in the inflationary throat.  For this discussion it
will not matter whether these are D3-branes or $\OL{\rm D3}$-branes, although
of course this will strongly affect the dynamics and the breaking of
supersymmetry.  With 3-branes in the throat, all the $(p,q)$ strings are
unstable to breakage.

The presence of the singularity gives rise to a new possibility.  We focus
on the example of the ${\bf C}^3/{\bf Z}_3$ orbifold singularity for
simplicity.  Hidden in the singularity are a collapsed two-cycle and a
collapsed four-cycle.  A D3-brane wrapped on the two-cycle and a D5-brane
wrapped on the four-cycle are both effectively one-dimensional, not only as
seen in four dimensions but as seen in ten --- they are fully located in the
throat.  Certain bound states of the D1 and wrapped D3 and D5 are referred to
as fractional strings, from their perturbative orbifold
description~\cite{fract,DG}.

If we consider the ${\bf Z}_3$ orbifold of the IIB theory before the ${\bf
Z}_2$ orientifold, and ignore for now the D3-branes, then the fractional
strings are BPS states.  They couple to the global $C_{(2)}$ field, and
also to two additional two-forms which are obtained from $C_{(4)}$
integrated over the collapsed two-cycle and from $C_{(6)}$ integrated over
the collapsed four-cycle.  In the perturbative picture the latter two
two-forms are twisted sector states.  The orientifold projection removes
$C_{(2)}$ but not the other two --- there are independent collapsed cycles
and corresponding two-form fields in each image throat, and one linear
combination of each survives.  Thus after the orientifold these fractional
strings are axionic.

To complete the story we must consider the $M^4$-filling branes.  As with
the D1-branes, there are fractional D3-branes, which arise from D5-branes
wrapped on the collapsed two-cycle and from D7-branes wrapped on the
collapsed four-cycle. The respective strings can break on these.  In the
models of ref.~\cite{AIQU} there are $M^4$-filling branes of all three types
--- there are three $U(1)$'s in particular --- so that all the strings are
unstable to breakage.

We conclude that with the Standard Model 3-branes in the inflationary throat
there are no interesting cosmic strings.

\section{Large dimension models}

In the \klmt model the string scale is lowered by a large warp factor.  It can
also be lowered in the context of large compact dimensions without such a
large warping; this was the context for the analysis in ref.~\cite{tye}.
Although in these models there are not yet examples with all moduli
stabilized, we can still investigate the stability of potential cosmic strings.

Consider first the prototype example, the $T^6/{\bf Z_2}$ orientifold, where
the ${\bf Z_2}$ reflects $k$ coordinates.  This model is equivalent to the type
I string under $T$-duality on $k$ axes.  It has D$p$-branes and O$p$-planes
for $p = 9-k$.  There are potential cosmic strings from wrapped D$q$-branes,
where $p-q$ must be even ($p-q$ odd gives non-BPS strings that immediately
decay~\cite{nonbps,ktheory}).  The key property that characterizes the
strings is the number $\nu$ of ND or DN directions, in which the
D$p$-branes are extended and the D$q$-branes are not or vice versa.  This number must be even (like $p-q$) and
is at least two, from the noncompact dimensions transverse to the
string.  If  $\nu = 2$ mod 4, the strings are non-BPS, like the D1-branes
of the \klmt model.  However, the strings here are free to fluctuate over
the whole compact space and so can rapidly find their ${\bf Z}_2$ images
and decay.  If $\nu = 4$ mod 4, the strings are BPS and so axionic; only
the $\nu = 2$ strings can break so these axions are not Higgsed.  Because this theory has ${\cal N}=4$ supersymmetry,
instantons cannot generate an axion potential, and the BPS strings are
stable global strings.

Reducing the supersymmetry to a realistic level will have several effects. 
First, it produces an axion potential so the BPS strings are now confined by
domain walls.  The instanton action is of order of the volume in string units of the cycle ${\cal K}_{7-p}$ on which it wraps.  If this is of order $10^{2}$ or more, the domain wall tension could be small enough not to confine the strings.  One would expect that the supersymmetry breaking would also produce a
potential for the positions of the D$q$-branes in the compact space, so that
these will be trapped in local minima.  If a D$q$-brane and its $\OL{{\rm
D}q}$ image are in different minima then their annihilation may be suppressed
as in the \klmt model.  There the suppression was due to a deep warp factor,
and here it would be due to a large distance $l$ between the D$q$ and
$\OL{{\rm D}q}$. Estimating the tunneling action~(\ref{schwing}) for the 
example of D1-branes, we have $\sigma \sim l / g_{\rm s} \alpha'$ and $\rho \sim 1/g_{\rm s} \alpha'$, and so
\begin{equation}
B \sim l^2/g_{\rm s} \alpha'\ .
\end{equation}
This can be quite large for large $l$.  Similarly there can be long-lived F-strings, and so $(p,q)$
strings.   There could be an even richer spectrum of strings in the large
dimension models from wrapping on different cycles~\cite{tye}.  The Standard
Model branes may allow some of these to break, but others --- those wrapped
on different cycles from the Standard Model branes, for example --- will
remain stable.

\section{Observational consequences}

\subsection{String properties}

The evolution of string networks, and their signatures, depend on certain
physical properties of the strings: their tensions, their interaction with
nongravitational fields, and their intercommutation properties.

\subsubsection{Tension}

For any given geometry the form of the brane-antibrane potential is known,
and so one obtains a relation between the observed magnitude of
density fluctuations $\delta_H$ and the parameters of the model.  For the
models of ref.~\cite{tye}, which are based on unwarped compactifications
with the moduli fixed by hand, the authors find a range
\begin{equation}
10^{-11} \lsim G\mu \lsim 10^{-6}\ ,
\end{equation}
with a narrower range around $10^{-7}$ for their favored models based on
branes at small angles.  For the \klmt model, combining equations~(3.7),
(3.9), and (C.12) of ref.~\cite{KKLMMT} gives
\begin{equation}
\frac{G^2 e^{4A_0}}{(2\pi\alpha')^2 g_{\rm s}} = \frac{\delta_H^3}{32\pi C_1^3
N_e^{5/2}}\ ,
\end{equation}
where $C_1$ is a model-dependent constant of order unity and $N_e$ is the
number of $e$-foldings for the observed fluctuations.  Inserting the
numerical values~\cite{KKLMMT} $\delta_H = 1.9 \times 10^{-5}$, $C_1 =
0.39$, and $N_e = 60$ gives $4 \times 10^{-20}$ for the right-hand
side.  The left-hand side is simply the product of the values of $G\mu$ for
the F-string and the D-string, assuming for simplicity that the RR scalar
$C$ vanishes.  Thus,
\begin{equation}
\sqrt{G\mu_{\rm F}\, G \mu_{\rm D}} \sim 2 \times 10^{-10}\ ,\qquad
\frac{\mu_{\rm D} }{ \mu_{\rm F} } = \frac{1}{g_{\rm s}} \ .
\end{equation} 
The value of $g_{\rm s}$ is likely in the range $0.1$ to
$1$ ($g_{\rm s} > 1$ can be converted to $g_{\rm s} < 1$ by $S$-duality),
so
$G\mu$ for both strings is in the range $10^{-10}$ to $10^{-9}$.

Note that these string tensions are all deduced assuming that the
cosmological perturbations arise from the fluctuations of the inflaton
field.  More general mechanisms would allow a wider range of scales.

\subsubsection{Superconductivity}

Superconducting strings carry massless charged degrees of
freedom~\cite{wsup}. Of the strings that we have discussed, the $(p,q)$
strings are not superconducting.  They are at a potential minimum distant
from any other branes in the compact space, and so their only massless
fluctuations are their transverse fluctuations and the fermionic partners of
these.  

The D-strings, obtained when a D7-brane passes through the throat, will be
superconducting with respect to the D7-brane gauge field if the D7-brane and
the D1-brane are precisely coincident.  In general there is no reason that
this should be the case because the positions are determined by different
dynamics --- that of the D1-brane by the local warp factor, and that of the
D7-brane by global considerations.  Further, the virtual 1-7 strings
contribute a short-distance repulsive term.  Thus these are unlikely to be
superconducting.   If the D1-D7 separation is small there will be a strong
coupling of the string to the Standard Model fields; such a coupling is
absent for the $(p,q)$ strings. 

\subsubsection{Intercommutation}

Consider first the case that there is only one kind of string.  When two
strings cross they may pass through one another or intercommute
(reconnect) as in figure~4.
\EPSFIGURE[h]
{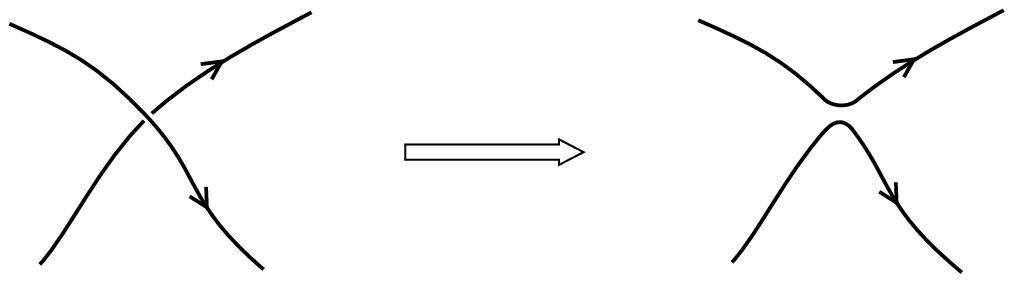}
{When two strings of the same type cross they can intercommute
(reconnect).}
For gauge theory strings the reconnection probability is essentially
unity~\cite{matz}.  For fundamental bosonic strings the probability was
obtained in ref.~\cite{coll}.  The same
result holds for the superstring~\cite{JJP}: the probability is of order
\begin{equation}
P \sim 0.5 {g_{\rm s}^2} \frac{V_{\rm min}}{V}
\end{equation}
for a typical collision.  Here $V$ is the
six-volume within which the strings are confined, and $V_{\rm min} = (4\pi^2
\alpha')^3$ is the minimum volume for toroidal compactification.  For
$g_{\rm s} > 1$ (and so also for D-strings) $P$ should saturate, possibly
with a slow rise, around $P \sim {V_{\rm min}}/{V}$.

For the large dimension models, the factor
$V_{\rm min}/{V}$ will give a large suppression if the strings are free to fluctuate over the compact space~\cite{tye}.  This may be the case in highly supersymmetric models, but as we have noted in
the previous section we expect that in realistic models any strings will be localized by a
potential and so the effective $V$ will be smaller than the total volume of
the compact space.  For a string of tension $\bar \mu$ whose transverse
fluctuations feel an external potential characterized by mass-squared $\OL m^2$ (the bars indicate that we are
working in the ten--dimensional metric), the quantum fluctuations of the
string give 
\begin{equation}
\langle (\delta X)^2 \rangle \sim \frac{1}{4\pi\bar\mu} \ln
\frac{\bar\mu}{\OL m^2}
\sim V_{\rm min}^{1/3} \frac{\bar \mu_{\rm F}}{8\pi^2\bar \mu} \ln 
\frac{\bar\mu}{\OL m^2} \ .
\end{equation}
We have used two-dimensional free field theory with a UV cutoff at $\bar\mu$.
From this it appears that the fluctuations around the potential minimum will
not give a large enhancement of $V/{V_{\rm min}}$; in particular, the effective $V$ grows only logarithmically with $m^{-1}$.   Hence the probability (5.4) for strings to intercommute is not unusually suppressed.  Thermal
fluctuations at $T \gg m$ would give some suppression of $P$.

When strings of two different types cross they cannot intercommute in the
same way.  Rather they can produce a pair of trilinear vertices connected
by a segment of string (figure 5).
\EPSFIGURE[h]
{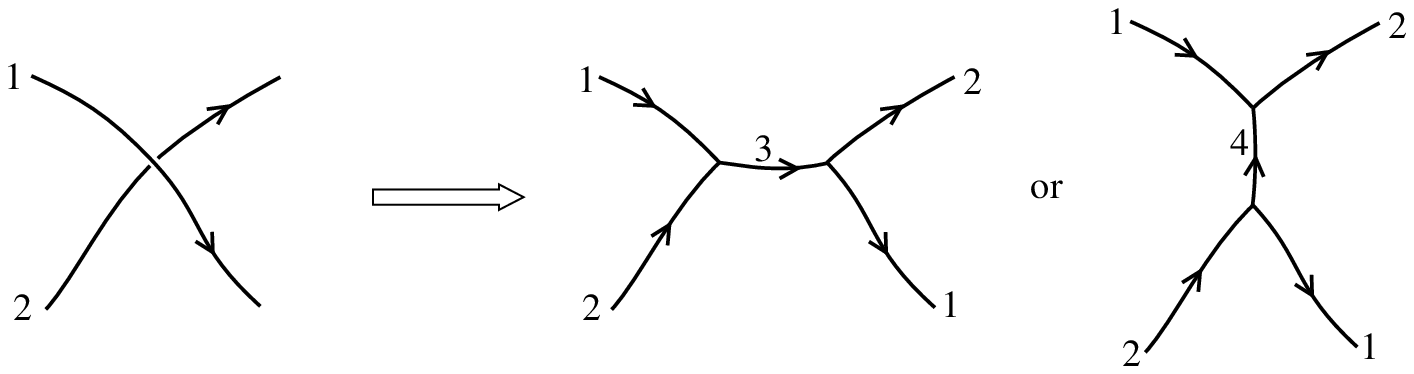}
{Crossing of strings of different types.}
For example, the crossing of a $(p,q)$ string and a $(p',q')$ string can
produce a $(p+p',q+q')$ string or a $(p-p',q-q')$ string.\footnote{There has
been extensive study of two-dimensional supersymmetric networks formed in
this way; see for example ref.~\cite{bpsnet}.}  Only for $(p',q') = \pm
(p,q)$ is the usual intercommutation possible.

A more complete treatment of this subject is in preparation~\cite{JJP}.

\subsubsection{Network properties}

After their formation at the end of inflation, string networks decay by the
combined processes of intercommutation, which breaks longer strings into
smaller loops, and gravitational radiation. For gauge theory strings there
is also the decay process due to the fields themselves, and
there is a debate as to which of the decay processes dominate
(gravitational versus Higgs) \cite{Vincent:1997cx,Moore:2001px}. Assuming
gravitational radiation dominates, then if the decay proceeds at the
maximum rate consistent with causality, the distribution of strings will
scale with the horizon volume.  This {\it scaling behavior} leads to an
energy density that goes as $t^{-2}$, so that $w = 1/3$ in the
radiation-dominated era and $w=0$ in the matter-dominated era.  That is,
in each era the energy density in strings is a fixed fraction of the
total energy density.  Simulations indicate that networks composed of a
single kind of string do scale, with  $\rho_{\rm string}/\rho_{\rm rad}
\sim 400 G \mu$ in the radiation dominated era and
$ \rho_{\rm string}/\rho_{\rm mat} \sim 60 G
\mu$ in the matter dominated era.  These values are for recombination
probability $P=1$ and increase as $P$ decreases~\cite{kib}.  

When there are several kinds of string, with trilinear vertices, then there
is the possibility that the network evolves to a three-dimensional structure
which freezes in a local minimum of the potential energy and then just grows
with the expansion of the universe.  In this case the energy density would
evolve as $a(t)^{-2}$, or $w= -1/3$.  Whether the network freezes or scales
is a complicated dynamical problem.  Such networks arise in field theory
when a symmetry group $G$ breaks to a discrete subgroup $K$. When $K$
is non-Abelian, intercommutation cannot occur, rather the network evolves
as in figure~5.  Simulations of $K=Z_3$ (with string vertices provided by
monopoles)~\cite{VV} and $K=S_3$~\cite{SP,McG} indicate that these systems
scale rather than freeze, but with some enhanced density of strings. 
Simulations of $K = S_8$~\cite{SP} show an energy density that grows
relative to the scaling solution and appears to indicate freezing
behavior.  This is consistent with the fact that the larger group allows
networks of greater topological complexity, but it could also be a
reflection that the simulations in~\cite{SP} have not yet managed to reach
the scaling regime. It is worth investigating this issue further.

To determine whether networks of
$(p,q)$ strings scale or freeze will ultimately require simulations.  We
conjecture that they scale, in that their topological complexity appears to
be roughly that of the
$S_3$ networks.  If $g_{\rm s}$ is close to one, only the four lowlying
strings with
$|p|,|q|\leq 1$ are likely to be heavily populated.  For any crossing
between two lowlying strings, one of the two processes in figure~5 will
again involve only lowlying strings, and for most angles of crossing this
process will be energetically favored. Another argument for scaling
behavior is to consider the limit
$g_{\rm s} \ll 1$, where the D-strings are much heavier than the F-strings. 
The D-strings should then evolve largely independently of the F-strings,
and so scale like a single-string network; after the D-strings decay to the
scaling distribution on a given length scale, the F-strings in turn evolve
like a single-string network.

\subsection{Observational bounds}

If a string network freezes into a $w=-1/3$ state, it quickly comes to
dominate the energy density of the universe unless the initial energy scale
is much lower than those considered above: it must be of order the weak
scale or less.  Thus if $(p,q)$ string networks freeze, models with 
$(p,q)$ strings are excluded with the parameters considered here.  Again,
our conjecture is that they do not freeze.

Assuming a scaling distribution of cosmic strings, the current upper bound
on
$G\mu$ comes from the power spectrum of the CMB, based on numerical
evolution of the Nambu-Goto equations: $G\mu
\lsim 0.7\times 10^{-6}$~\cite{Landriau:2003xf} (see also
ref.~\cite{tye2}).  The tensions given in section 5.1.1 for the various
models are below this bound.  On the other hand numerical evolution of the
underlying Abelian-Higgs field theory has led  Vincent et al to argue that the
bound is closer to
$G\mu
\approx 10^{-8}$~\cite{Vincent:1997cx} (however see
also~\cite{Moore:2001px}). 

Superconducting strings may act as sources for
vortons~\cite{Davis:ij}, loops of cosmic string with charge and current
stabilized by the angular momentum of the charge carriers.  In this case
they would be subject to bounds on their allowed tension,  with $10^{-28}\lsim
G\mu \lsim 10^{-10}$ being claimed to be a cosmologically unacceptable range of
values~\cite{Martins:1998th}.  If the energy scale
associated with superconducting strings were close to the electroweak scale,
then the vortons could become serious candidates for cold dark matter. In the
context of the \klmt model, this would correspond to having inflation in the
throat occurring at or around the electroweak scale.

Cosmic strings produce large quantities of
gravitational waves, because they are relativistic and inhomogeneous. 
Pulsar timing measurements then place an upper bound on $G\mu$ which is
roughly comparable to that from the CMB, depending on uncertainties from
network properties \cite{CBS}.

Remarkably, future measurements of non-gaussian emission of 
gravitational waves from cusps on strings
will be sensitive to cosmic strings with values of $G\mu$ seven orders of
magnitude below the current bound, covering the entire range of
tensions discussed in section 5.1.1.  According to ref.~\cite{DV}, even LIGO 1
may be sensitive to a range around $G\mu \sim 10^{-10}$, while LIGO 2 will
reach down to $G\mu \sim 10^{-11}$ and LISA to $G\mu \sim 10^{-13}$.  In
addition~\cite{DV}, pulsar timing measurements may reach a sensitivity of
$G\mu \sim 10^{-11}$.  Thus, gravitational waves provide a potentially large
window into string physics, if we have a model in which strings are produced
after inflation and are metastable.

If the string couples strongly to Standard Model fields then instead of
primarily  producing gravitational radiation the string network may decay
through the production of high energy cosmic rays, photons and neutrinos
from string cusps~\cite{Wichoski:1998kh}.   These authors have calculated
the predicted flux of high energy gamma rays, neutrinos and cosmic ray
antiprotons and protons as a function of the scale of symmetry breaking at
which the strings are produced, and argued that in order to reproduce the (possibly)
observed distribution of particles above the GZK cut-off, they require $G\mu
\leq 10^{-9}$. Given the values we expect in the \klmt model this
remains in the interesting regime for cosmic strings arising out of string
theory.  Note however refs.~\cite{BPO}, which argue that the cosmic radiation from cusps is suppressed.

\section{Conclusions}

We have found that both fundamental and Dirichlet strings might be observed
as cosmic strings.  The issue is model-dependent --- it depends on having
brane inflation to produce the strings, and on having a scenario in which
the strings are stable.  Nevertheless, this is a potentially large and
rather direct window onto string theory. 

Of course, if cosmic strings are discovered, the problem will be to
distinguish fundamental objects from gauge theory solitons.  Indeed, this is
not a completely sharp question, because these are dual descriptions of the
same objects.  If one can infer that the strings have intercommutation
probabilities less than unity, this is a strong indication that they are
weakly coupled F-strings.  Discovery of a
$(p,q)$ spectrum of strings would be a promising signal for F- and
D-strings.  Note however that these throats have a dual gauge
description~\cite{KS} and therefore such strings can also be obtained in
gauge theory; the spectrum is actually a signal of an $SL(2,{\bf Z})$
duality and so might arise in other ways as well.  If cosmic strings are
found through the gravitational radiation from cusps, determining their
tensions and intercommutation properties will require a spectrum of many
events as well as precise simulations of the evolution of string networks.

\acknowledgments

We would like to thank Stephon Alexander, David Berenstein, Robert
Brandenberger, Alessandra Buonanno, Thibault Damour, Gia Dvali, Jaume Gomis, Chris
Herzog, Nick Jones, Shamit Kachru, Renata Kallosh, Tom Kibble, Andrei
Linde, Eva Silverstein, Scott Thomas, Sandip Trivedi, Mark Trodden, Henry Tye, and
Alex Vilenkin for useful discussions.  We also thank Jose Blanco-Pillado, Greg Moore, and Ken Olum for comments on the manuscript.  EJC and RCM would like to thank
the organizers of the String Cosmology program at the Kavli Institute for
Theoretical Physics for their invitation to participate in such a
stimulating meeting.  The work of RCM at the Perimeter Institute is
supported by funds from NSERC of Canada. The work of JP
is supported by National Science Foundation grants PHY99-07949 and
PHY00-98395.

\end{document}